\documentclass[11pt,a4paper]{article}
\usepackage{subfigure}
\usepackage{xcolor}
\usepackage[T1]{fontenc} 
\usepackage{xspace}
\usepackage{multirow}
\usepackage{hyperref}
\usepackage{slashed}

\usepackage{graphicx,amsmath}
 \usepackage{amssymb}
\usepackage{lineno,color}
\usepackage{url}

\def\be{\begin{equation}}                                                    
\def\ee{\end{equation}}                                                    
\begin{document}
 

\bigskip
\bigskip
\begin{center}
{\bf\Large Current Status of the Odderon~\footnote{Talk given at the Conference ``Hadron Structure and Fundamental Interactions:
from Low to High Energies",  Gatchina, Russia, July 8 – 12, 2024 and XXXVI
International Workshop on High Energy Physics “Strong Interactions: Experiment, Theory, Phenomenology”, Protvino, Russia, July 23-25, 2024.} }\\
 
 \vspace{1cm}

 M.~G.~Ryskin\\
 
 \vspace{0.7cm}
 Petersburg Nuclear Physics Institute, NRC Kurchatov Institute, \\Gatchina, St.~Petersburg, 188300, Russia\\
\
 \vspace{0.7cm}

 \abstract{\noindent Odderon is the C-odd amplitude which does not die out (or die very slowly) with energy.  We consider the constrains on the Odderon properties and the
perturbative QCD odderon given at the lowest $\alpha_s$ order by the three gluon exchange. Then we discuss the experimental indications for the odderon
contribution to high energy proton-proton elastic scattering and some other processes in which the odderon may reveal itself.}\\
  \vspace{1cm}

 \vfill

 E-mail:  \url{ryskin@thd.pnpi.spb.ru}
 
\end{center}
 \newpage
\section{Introduction}
Odderon is the C-odd amplitude which does not die out (or die very slowly) with energy. Theoretically there are no reasons to have {\em no} such an amplitude. Moreover, we have it in perturbative QCD where at the lowest $\alpha_s$ order it is given by the three gluon exchange when all three gluons are symmetric in colour (i.e. convoluted by the colour SU(3) tensor $d^{abc}$).\\

\begin{figure} [t]
\label{f1}
\begin{center}
\vspace{-2cm}
\hspace{-0.6cm}
 \includegraphics[scale=0.36]{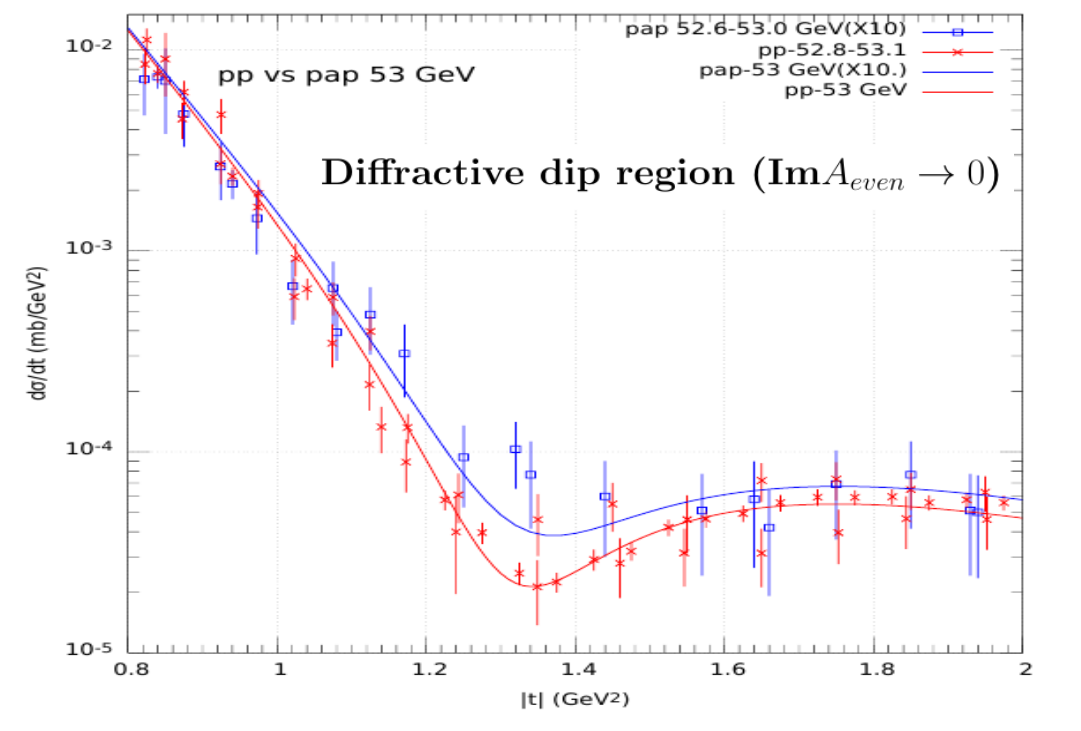}
\vspace{-.4cm}
\caption{\small Differential cross section $d\sigma/dt$ for
$pp$ (red) and $\bar pp$ (blue) elastic scattering at $\sqrt s=53$ GeV. Data are from \cite{53}.}
\end{center}
\end{figure}

First time the odderon was discovered in 1985 when the elastic $pp$ and $\bar pp$ scattering were measured at CERN-ISR at $\sqrt s=53$ GeV and the difference between $pp$ and $\bar pp$  cross sections was observed in diffractive dip region (see Fig.\ref{f1}). However the statistic was not sufficiently large and the energy was not high enough to completely neglect the C-odd secondary Reggeon contributions.

Since the Odderon amplitude is expected to be rather small the best chance to observe it on the top of a much larger C-even contribution is either in diffractive dip region where the imaginary part of C-even amplitude vanishes or looking for the real part of $pp$ ($\bar pp$) elastic amplitude. Recall that due to dispersion relations the real part of high energy C-even amplitude is relatively small (Re$A_{even}<<$Im$A_{even}$).\\

The real part of proton-proton amplitude can be measured  
via the Coulomb-nuclear interference (CNI) at very low momentum transferred $|t|$. In 2018-2020 TOTEM claimed the Odderon discovery based on two results. First, they  measured the elastic $pp$ scattering at low $|t|$ down to $-t=8\cdot 10^{-4}$ GeV$^2$ and determine the real to imaginary part ratio $\rho=$Re/Im$\simeq 0.09-0.10$ which turns out to be~\cite{rho-T}
 about 0.04 smaller than that ($\sim 0.13-0.14$) coming from the dispersion relations for a pure C-even scattering.
 
 Next the $pp$ cross section in  diffractive dip region was measured at 2.76 GeV \cite{2.76}. A clear 'dip-bump' structure was observed while at a relatively close Tevatron energy  of 1.96 TeV the $t$ dependence of $\bar pp$ cross section is more or less flat \cite{D0-T}. This can be explained by the presence of the Odderon real part which diminishes the real part of $pp$ amplitude but enlarges it filling the dip in $\bar pp$ case. 

In section 2 we consider the unitarity constraints on high energy C-odd amplitude and the lowest order QCD expectations.

In section 3 we discuss in more details the situation in CNI region and the new ATLAS-ALFA 13 TeV results.

Then in sections 4 and 5 the diffractve dip region and some other processes in which the Odderon contribution can be observed will be discussed shortly.

We conclude in section 5.\\

 \section{Theory}
The major constraint on C-odd amplitude comes from the fact that both the particle and the antiparticle cross sections should be positive while the C-odd amplitude changes its sign going from particle to antiparticle.
 This condition must  be fulfilled at any energy and at each impact parameter, $b$, that is at any partial wave $l=b\sqrt s/2$.
 
 This means that the intercept 
 \be 
 \label{1}
 \alpha_{Odd}(0)<\alpha_{even}(0)\quad\mbox{and the slope}\quad  B_{Odd}<B_{even}\ .
 \ee
The pert.QCD satisfies these conditions. For the lowest $\alpha_s$ order 3 gluon diagram we get $\alpha_{Odd}(0)=1$ while the Pomeron intercept $\alpha_{Pom}(0)>1$.

In the leading (in $\ln(s)$) order the Odderon intercept is equal to 1~\cite{Lip} or a bit smaller than 1 ($\alpha_{Odd}=0.96-1$, see \cite{BE}).

Besides this it was shown that in  $b$ space the QCD Odderon occupies the area of a smaller radius, see e.g. \cite{BGV}.

Note that in QCD the C-odd amplitude is expected to be smaller than the C-even one. First, it is proportional to $\alpha_s^3$ and not $\alpha_s^2$ as in C-even case. Next, while the BFKL Pomeron contribution is driven by the {\em maximum} quark-quark separation 
\be
A_{even}(t=0)\propto \alpha^2_s<r^2_{max}>
\ee
for the Odderon we expect \cite{FK,Ro}
\be
A_{Odd}(t=0)\propto \alpha_s^2<r^2_{min}>\ .
\ee
Recall that the Odderon does not couple to pion since the C-parity of $\pi^0$ is positive. Describing  the proton by ``quark-diquark" model we get  the same colour structure as that in pion. That is for a point-like diquark the proton-Odderon coupling, $\beta_O(t)$, should vanishes at $t=0$\;\footnote{At $t\neq 0$ we get non zero result, $\beta_O\neq 0$, due to a larger diquark mass (see \cite{Z} for more details)}. This means that at $t=0$ the Odderon amplitude is proportional to  the separation of quarks inside the diquark, that is to the minimal value of $<r^2>=<r^2_{min}>$.
\section{C-odd contribution to real part of   $A(t=0)$} 
The real part of C-even amplitude can be calculated based on dispersion relations. At high energies it takes the form
\be
\label{disp}
\rho_{even}\simeq \frac{\pi}{2}\frac{\partial\ln \sigma_{tot}(s)}{\partial\ln s} \ .
\ee
That is the value of $\rho$ is strongly correlated with the energy behavior of total cross section. Using the cross sections measured by TOTEM without the Odderon we expect $\rho=\rho_{even}=0.13-0.14$ and not $\rho=0.09-0.10$ observed by TOTEM \cite{rho-T}. However the normalization used by TOTEM is questionable. ATLAS systematically publish a smaller $\sigma_{tot}$. Therefore it is better to describe  data introducing the normalization parameter, $N$, for each set of measurements. Of course the deviation of $N$ from 1, divided by the corresponding error in luminosity, is included in the total $\chi^2$ value.

Petrov and Tkachenko fitted the TOTEM 13 TeV data \cite{rho-T} accounting for the correlated errors and including the normalization factor as the free parameter. That is actually the normalization to the Coulomb low $t$ scattering was used. They obtained a smaller $\sigma_{tot}=107.6\pm 1.7$ mb 
 and a bit larger $\rho=0.11\pm 0..01$.\\
 
 New ATLAS-ALFA 13 TeV data \cite{atl13} confirm the TOTEM $\rho=0.10\pm 0.01$ but gives an even smaller cross section $\sigma_{tot}=104.7\pm 1.1$ mb.
 
 In the recent paper \cite{LRK} the available low $|t|<0.1$ GeV$^2$ data at $50\ \mbox{GeV}<\sqrt s<13$ TeV were analyzed
 including both the TOTEM and ATLAS-ALFA results with the corresponding (free) normalization factors.
 
 Two channel eikonal model
 $$ A(s,t) ~=~ is \int^{\infty}_{0} b\, db\, J_{0}(bq) \cdot$$
 \be
 \label{2eik}
 \left[ 1 -\frac{1}{4}\, e^{i(1+\gamma)^{2}\Omega(s,b)/2} \right. \nonumber \\
 \left. -\frac{1}{2}\, e^{i(1-\gamma^{2})\Omega(s,b)/2} - \frac{1}{4}\, e^{i(1-\gamma)^{2}\Omega(s,b)/2} \right]
 \ee
was used where the opacity $\Omega(s,b)=\Omega_{Pomeron}(s,b)+\Omega_{Odd}(s,b)$ is given by  sum of the even/Pomeron and the Odd terms.\\

We obtain a quite satisfactory fit with $\chi^2=560$ for 504
degrees of freedom, $\nu$; $\chi^2/\nu=1.11$.
 Neglecting the Odderon we get a much larger $\chi^2=726$. The quality of the Coulomb-nuclear interference region description is shown in Fig.2.\\
 \begin{figure} [t]
\label{ff2}
\begin{center}
\vspace{-2cm}
\hspace{-0.6cm}
 \includegraphics[scale=0.45]{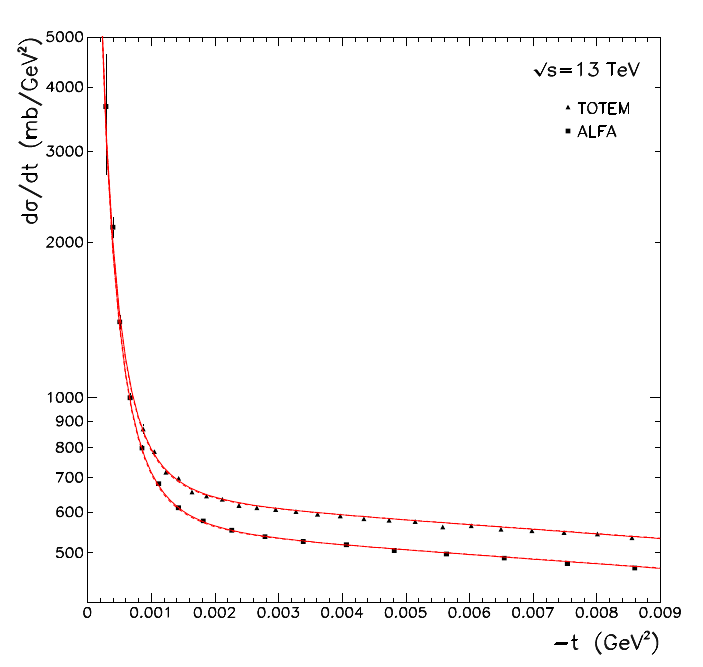}
\vspace{-.4cm}
\caption{\small Description of $t$ dependence of elastic $pp$ differential cross section in Coulomb-nuclear interference region. Data are from \cite{rho-T,atl13}. Theoretical curves are multiplied by the corresponding normalization factors.}
\end{center}
\end{figure}

The main lessons about the Odderon coming from this study are:
\begin{itemize}
\item The description using the Odderon improves the fit (with the Odderon  $\chi^2$ becomes much smaller).
\item The sign of the Odderon amplitude needed to describe the very low $|t|$ data is opposite to that predicted by the perturbative QCD three-gluon exchange contribution.
\footnote{The problem can be solved assuming that the Odderon coupling $\beta_O$ vanishes (or strongly decreases) at $t=0$. In this case the dominant C-odd contribution at $t=0$ comes from the Pomeron-Odderon cut and has another sign. }
 \item The Odderon-proton coupling, $\beta_O$, is smaller than that for the Pomeron. Moreover after via the eikonal we account for the screening of seed Odderon by the Pomeron the final $C$-odd 
  contribution to $\rho$ at 13 TeV becomes quite small,
$\delta\rho=(\rho^{\bar pp}-\rho^{pp})/2\leq 0.004$ -- i.e. 10 times smaller than that ($\delta\rho=0.04$) originally claimed by TOTEM. 
\end{itemize}

\section{Dip region}
In order to observe the Odderon in the diffractive dip region we have to measure both the $pp$ and $\bar pp$ cross sections at the {\em same} energy. At the moment we have no such  data. Therefore authors extrapolate the LHC $pp$ cross sections to a lower, Tevatron, $\bar pp$ energy. See for example
 \cite{D0-T} and the recent paper \cite{CNPSS}. Unfortunately we have no solid theory for this extrapolation and therefore any inaccuracy in extrapolation may looks as the observation of the Odderon contribution. 

\section{Other processes driven by the Odderon}
At first sight the best reaction in which the Odderon exchange dominates
 is the C-even meson photoproduction. It was proposed in \cite{Na} to search for the Odderon at HERA observing the exclusive $\pi^0$ production $\gamma+p\to \pi^0 +p$ but only the upper limit to corresponding cross section was obtained. 
 The problem is a large background coming from the vector meson production $\gamma+p\to V+p$ with the subsequent radiative decay, say, $\omega\to \pi^0+\gamma$.
 
 Such background can be suppressed in future Electron Ion Collider if the energy of the outgoing electron will be measured and the incoming photon energy will be compared with the C-even meson energy.
  
The possibility to searching for the Odderon in Ultraperipheral Proton--Ion Collisions at the LHC in exclusive   C-even meson production was discussed in \cite{HKMR}.\\

Another interesting process where the Odderon contribution may be important is the $K_L\to K_S$ regeneration. However it will be challenging to select the pure exclusive events while in not exclusive case the background caused by the triple Regge $\omega\omega\mbox{-Pomeron}$ and other similar diagrams looks too large.
\section{Conclusion} 
\begin{itemize}
\item Odderon exits in perturbative QCD and was observed
(at least the serious hints in favour of Odderon) experimentally.\\
I would not like to discuss the confidence level of Odderon discovery since the major uncertainties are not statistics but comes from systematics.
\item Odderon can be studied looking for the real part of high energy elastic scattering amplitude at $t\to 0$ or in diffractive dip region. It would be the best to measure and to compare the $pp$ and $\bar pp$ scattering at the {\em same} energy and with the {\em same} detector (may be at the lHC at 900 GeV).
\item
The Odderon-proton coupling is small.
\item 
Besides the elastic $pp$ and $\bar pp$ scattering another possibility to observe the Odderon is the C-even meson photoproduction and/or the $K_L\to K_S$ regeneration. 
\end{itemize}

 At present the goal is not to proof that the Odderon exists
(no reason to have {\em  No} Odderon)
 but to {\em measure} the Odderon exchange amplitude, at least at $t=0$ and in the dip region.\\
 
Maximum Odderon ($A_{Odd}\propto \ln^2s$) is another story. Taking s- and t- channel unitarities together we see that asymptotically the maximum Odderon contradicts unitarity (see~\cite{BD}) however this does not mean  that one can not use the maximum Odderon parameterization within  some limited energy range.

  \section*{Acknowledgments}  
  Author thanks V.A. Khoze for the discussion and for the reading  of this manuscript.

  

\end{document}